\input harvmac

\Title{\vbox{\baselineskip12pt
\hbox{BCCUNY-HEP/01-02} \hbox{hep-th/0109041}}}
{\vbox{\centerline{Fundamental Strings and Cosmology}}}

\baselineskip=12pt \centerline {Ramzi R.
Khuri\footnote{$^*$}{e-mail: khuri@gursey.baruch.cuny.edu.
Research supported by a Eugene Lang Junior Faculty Research Fellowship.}}
\medskip
\centerline{\sl Department of Natural Sciences, Baruch College,
CUNY} \centerline{\sl 17 Lexington Avenue, New York, NY 10010
\footnote{$^\dagger$}{Permanent address.}}
\medskip
\centerline{\sl Graduate School and University Center, CUNY}
\centerline{\sl 365 5th Avenue, New York, NY 10036}
\medskip
\centerline{\sl Center for Advanced Mathematical Sciences}
\centerline{\sl American University of Beirut, Beirut, Lebanon
\footnote{$^{**}$}{Associate member.}}

\bigskip
\centerline{\bf Abstract}
\medskip
\baselineskip = 20pt

We show that the velocity-dependent forces between parallel
fundamental strings moving apart in a $D$-dimensional spacetime
imply an expanding universe in $D-1$-dimensional spacetime.

\Date{September 2001}

\def\D{h}
\def\dx{\dot x}
\def\ddx{\ddot x}

\def\r{\rho}

\def\({\left (}
\def\){\right )}
\def\[{\left [}
\def\]{\right ]}

\lref\prep{M. J. Duff, R. R. Khuri and J. X. Lu, Phys. Rep.
{\bf B259} (1995) 213, hep-th/9412184.}

\lref\dab{A. Dabholkar, G. W. Gibbons, J. A. Harvey and F. Ruiz Ruiz,
Nucl. Phys. {\bf B340} (1990) 33.}

\lref\calk{C. G. Callan and R. R. Khuri, Phys. Lett. {\bf B261}
(1991) 363.}

\lref\astro{A. G. Reiss {\it et al.}, Astron. J. {\bf 116}
(1998) 1009, astro-ph/9805201; S. Perlmutter {\it et al.},
Astrophys. J. {\bf 517} (1999) 565, astro-ph/9812133.}

\lref\stcosm{See S. M. Carroll, hep-th/0011110; A. Linde, hep-th/0107176
and references therein.}

\lref\bps{M. K. Prasad and C. M. Sommerfield, Phys. Rev. Lett. {\bf 35}
(1975) 760; E. B. Bogomol'nyi, Sov. J. Nucl. Phys. {\bf 24} (1976) 449.}

\lref\dynam{R. R. Khuri and R. C. Myers, Phys. Rev. {\bf D52} (1995) 6988,
hep-th/9508045.}

\lref\rs{L. J. Randall and R. Sundrum, Phys. Rev. Lett.
{\bf 83} (1999) 3370;  Phys. Rev. Lett. {\bf 83} (1999)
4690.}

\lref\elias{A. Kehagias and E. Kiritsis, JHEP {\bf 9911} (1999) 022.}



In this paper, we show that the velocity-dependent forces between
parallel fundamental strings in $D$ spacetime dimensions, with certain initial
conditions, lead to an expanding universe in $D-1$ dimensions. While the model
presented below is very simple, the result seems to be quite general and is  expected to hold for more complicated scenarios. The findings seem consistent
with recent observations \astro\ of an accelerating universe, with predictions
for the late time expansion rate.

We start with the action $S=I_D + S_2$, where
\eqn\sgact{I_D = {1\over 2\kappa^2} \int d^D x \sqrt{-g}
e^{-2\phi} \left(R+ 4(\partial\phi)^2 -{1\over 12} H_3^2\right)}
is the $D$-dimensional string low-energy effective spacetime action and
\eqn\smact{S_2 =-{\mu\over 2} \int
d^2 \zeta \left(\sqrt{-\gamma} \gamma^{\mu\nu} \partial_\mu X^M
\partial_\nu X^N g_{MN} +\epsilon^{\mu\nu} \partial_\mu X^M
\partial_\nu X^N B_{MN}\right)}
is the two-dimesional worldsheet
sigma-model action. Here $g_{MN}$, $B_{MN}$ and $\phi$ are the
spacetime sigma-model metric, antisymmetric tensor and dilaton,
respectively, while $\gamma_{\mu\nu}$ is the worldsheet metric.
$H_3 = dB_2$ and $\mu$ is the string tension.

The ``fundamental string'' (or ``elementary'' string) solution to
the combined action, representing stationary macroscopic strings
parallel to the $x^1$ direction, is given by \dab
\eqn\fstring{\eqalign{ds^2 & = \D^{-1} \left(-dt^2+(dx^1)^2\right)
+ \delta_{ij} dx^i dx^j,\cr
e^{-2\phi} & = - B_{01} = \D = 1 + {k\over r^n},\cr}}
where $n=D-4$, $r^2 = x^i x_i$ and
the indices $i$ and $j$ run through the $D-2$-dimensional
space transverse to the string. The constant $k$ is essentially
the Noether charge of the source string and is proportional to its
tension.

This solution can be extended to a multi-static string solution
owing to the existence of a zero-force condition. This condition in turn
arises from the cancellation between the attractive gravitational and
dilatational forces of exchange with the repulsive antisymmetric field
exchange, and is based on the existence of supersymmetry and the saturation
of a BPS bound \bps.

It was subsequently shown that, in addition to the zero static force,
the leading order ($O(v^2)$) velocity-dependent forces cancel for moving strings
as well \calk\ (see also \prep).
This result too is associated with the existence of higher supersymmetry \dynam.

Consider the Lagrangian for a test fundamental string moving in
the background of another, source string. Replacing the fields of
\fstring\ in \smact\ leads to the Lagrandian
\eqn\lag{{\cal L}=-\mu \D^{-1} \left(\sqrt{1-\D \dx^2} - 1\right),}
where $\dx ^2 = \dx^i \dx_i$ and the ``$\cdot$'' represents a time derivative.
The Euler-Lagrange equations for \lag\ lead via a straightforward
calculation to
\eqn\euler{\ddx^i-{nk\dx^2 (x^j \dx_j) \dx^i \over r^{n+2} \left(1-\D \dx^2 \right)}
+ {\D (\dx^k \ddx_k)\dx^i \over \left(1-\D \dx^2 \right)}= {\partial_i \D\over \D^2}
\left(1-{\D \dx^2\over 2} -\sqrt{1-\D \dx^2} \right).}
Contracting with $\dx_i$ and simplifying leads to
\eqn\eulerc{\dx^i \ddx_i = {nk x^j \dx_j\over 2 \D^2
r^{n+2}} \left( (y-1)^2 (2y+1)\right),}
where $y = \sqrt{1-\D
\dx^2} \geq 0$. Since the term in parentheses in the right hand
side of \eulerc\ is always positive, it follows that whenever $x^j
\dx_j > 0$, it follows that ${d\over dt} (\dx^2)=2\dx^i \ddx_i > 0$.
In other words, whenever the component of the relative velocity of the
strings in the direction of their line of separation is parallel
(as opposed to antiparallel) to the relative position vector, the
relative speed is increasing. This means that the
velocity-dependent force is repulsive whenever the strings are
moving away from each other. This in turn has an interesting
consequence for the separation distance between the strings.

It is straightforward to show from the Hamiltonian $H = \dx^i p_i
- {\cal L}$ following from \lag\ that the conserved energy of the
two-string system is given by \eqn\energy{E={\mu \over \D}
\left({1\over \sqrt{1-\D \dx^2}} -1\right).} Setting $\dx^2=u$,
the function \eqn\eff{f(u,\D) = {1\over \D} \left({1\over
\sqrt{1-u\D}} - 1\right) = {E\over \mu}=\r} is a constant. It is
straightforward to show that $\partial f/\partial u > 0$ and
$\partial f/\partial \D > 0$ for all values of $u$ and $\D$ (note
that $u \leq 1$, $\D > 1$ and $u\D < 1$). Therefore, if $u =\dx^2$
increases, as is the case above, then $\D$ must decrease. This
follows from requiring $df= (\partial f/\partial u) du + (\partial
f/\partial \D) d\D=0$, so that if $du>0$, then the positivity of
the partial derivatives of $f(u,\D)$ implies that $d\D <0$. If $\D
= 1+k/r^n$ decreases, it follows that $r$, the separation of the
strings, must increase. Therefore if the strings are initially
moving apart, the net velocity-dependent force between them is
repulsive, which leads to a further separation of the strings.
Since this type of interaction occurs for any two strings, if we
start with any number of close, parallel strings initially moving
apart in the transverse space, they will continue
to do so indefinitely and will fuel an expanding universe in the
$D-2$-dimensional transverse space and therefore in the
$D-1$-dimensional spacetime orthogonal to the strings. For
example, five-dimensional fundamental strings lead to an expanding
universe in $D=4$ spacetime dimensions.

The time-dependence of the general expansion is also interesting.
From \eulerc, it is easy to see that the acceleration vector
$\ddx^i$ is always in the plane determined by the position vector
$x^i$ and the velocity vector $\dx^i$. It follows that the motion
of the test-string in the two-body problem remains in the same
plane. This allows us to use polar coordinates and write the
Lagrangian as \eqn\lagtb{{\cal L}=-\mu \D^{-1} \left(\sqrt{1-\D
\left( \dot r^2 + r^2 \dot\theta^2\right)} - 1\right).} This
Lagrangian has a conserved angular momentum \eqn\ang{L
={\partial {\cal L}\over \partial \dot\theta} = {\mu r^2
\dot\theta \over \sqrt{1-\D \left( \dot r^2 + r^2
\dot\theta^2\right)}}= l \mu.} Solving for $\dx^2$ from \energy,
one easily arrives at \eqn\velo{\dx^2 = \dot r^2 + r^2\dot
\theta^2= {\r (\D \r +2)\over (\D \r + 1)^2}.} Solving for
$\dot\theta^2$ from \ang\ and replacing in \velo\ leads to
\eqn\veloang{\dot r^2 = {\r (\D \r +2) \left( 1 +
l^2\D/r^2\right)\over (\D \r +1)^2 \left( 1 + l^2/r^2\right)}.}

Now consider a universe consisting of many fundamental strings
initially very close to each other. Each pair of such strings
interacts as above, so that an initial outward propagation of the
strings would tend to further push them apart in the transverse
space. The exact solution of this problem requires a many-body
computation, probably best done by numerical methods. For now, to
obtain a rough idea as to the time dependence of the expansion of
such a universe, let us assume that in a mean-field approximation,
the effective force on each string may be approximated by that of
a single, very large source fundamental string whose Noether
charge $k$ is equal to the total charge of all of the strings in
the $D$-dimensional space. In this approximation, the existence of
nonzero angular momentum $l$ for the test string depends on the
presence of an initial angular momentum in the universe. The
distance $r$ between the test string and the source string in this
model then represents the approximate average position of the
strings, and hence the size of the universe.  We wish to study the
time dependence of $r$ at both early and late times for an
expanding model.

In the very unlikely scenario of $l >> k^{1/n}$, an immediate
consequence of \veloang\ is that the conditions $\dot r \leq 1$
and $1-\dx^2 h > 0$ rule out this model for early times $r <<
k^{1/n}$. Furthermore, even with the more reasonable assumption of
$l << k^{1/n}$, the model breaks down in the domain $r << l$ for
the same reasons. It is therefore reasonable to assume $l\simeq 0$
in the mean-field approximation, so that the general expansion can
be regarded as essentially radial, {\it i.e.} the angular
components of the velocity-dependent forces average to zero and
$\dx^2 \simeq \dot r^2$. In any case, the existence of small
nonzero $l$ does not greatly affect the qualitative conclusions
regarding the expansion rate. In all domains, the strings are
separating at an accelerated rate, with both $\dot r > 0$ and
$\ddot r > 0$. For early times, $l << r << k^{1/n}$, with $r >>
(kl^2)^{1/(n+2)}$, the time-dependence of $r$ also depends on
$n=D-4$, and hence on the dimension of the spacetime. For $n>2$,
$r \sim (b-t)^{-1/(n/2 -1)}$, where $b$ is a constant. The case of
$n=2$ (or $D=6$) is the most interesting: $r \sim
\exp{(t/\sqrt{k})}$, representing an exponential inflationary
phase. For $n=1$, $r \sim t^2$. For all $n \geq 1$, the expansion
proceeds with positive, outward acceleration. At very late times,
for $r >> k^{1/n}$, the expansion rate asymptotically approaches a
constant speed $\dot r \to \sqrt{\r (\r +2)/(\r +1)^2}$, again for
all $n \geq 1$.

Without attaching too much importance to the exact quantitative
predictions of this highly simplified model, one can still extract
interesting consequences from this kind of string-seeded universe.
The main feature is that, for certain initial conditions, the
expansion is self-fueled by the repulsive velocity-dependent
forces. This occurs because the initially separating strings lead
to a repulsive force, which further separates the strings, etc.
The exponentially expanding phase for $D=6$ is interesting, and
suggests an inflationary phase.  An interesting possibility in
this case is that the moving strings in $D=6$ lead to an expanding
five-dimensional universe, in which an effective four-dimensional
brane universe resides, following \rs. The asymptotic late time
expansion rate of $\sqrt{\r (\r +2)/(\r +1)^2}$ is also
intriguing, and may represent a testable prediction for this type
of model. Further investigations of this type of model are clearly
merited, using a possible combination of numerical computations,
quantum string effects and nonequilibrium thermodynamics.

{\bf Note Added:} After this paper appeared, it was pointed out to the author
that a similar approach to string cosmology, though with somewhat
different focus, was used in \elias.

\listrefs
\end